\documentclass[10pt, a4paper,onecolumn]{article} 

\pdfoutput=1

\let\d\donothing

\newcommand{\bra}[1]{\langle #1|}
\newcommand{\ket}[1]{|#1\rangle}
\newcommand{\expt}[1]{\langle#1\rangle}

\newcommand{\d}[0]{\partial}

\title{Entropic Updating of Probability and Density Matrices}

\author{Kevin Vanslette}

\author{Kevin Vanslette\\ kvanslette@albany.edu\\
\\Department of Physics, University at Albany (SUNY)\\
Albany, NY 12222, USA}

\begin{document}
\maketitle
\abstract{We find that the standard relative entropy and the Umegaki entropy are designed for the purpose of inferentially updating probability and density matrices respectively. From the same set of inferentially guided design criteria, both of the previously stated entropies are derived in parallel. This formulates a quantum maximum entropy method for the purpose of inferring density matrices in the absence of complete information in Quantum Mechanics. }

\section{Introduction}
We \emph{design} an inferential updating procedure for probability distributions and density matrices such that inductive inferences may be made. The inferential updating tools found in this derivation take the form of the standard and quantum relative entropy functionals, and thus we find the functionals are \emph{designed} for the purpose of updating probability distributions and density matrices respectively. Design derivations which found the entropy to be a tool for inference originally required five \emph{design criteria} (DC) \cite{Shore1,Shore2,Csiszar1991}, this was reduced to four in \cite{Skilling1, Skilling2, Skilling3}, and then down to three in \cite{book}. We reduced the number of required DC down to two while also providing the first \emph{design} derivation of the quantum relative entropy -- \emph{using the same design criteria and inferential principles in both instances}.

The designed quantum relative entropy takes the form of Umegaki's quantum relative entropy, and thus it has the ``proper asymptotic form of the relative entropy in quantum (mechanics)" \cite{Petz1, Petz2, Petz3}. Recently, \cite{Wilming} gave an axiomatic characterization of the quantum relative entropy that ``uniquely determines the quantum relative entropy". Our derivation differs from their's, again in that we \emph{design} the quantum relative entropy for a purpose, but also that our DCs are imposed on what turns out to be the functional derivative of the quantum relative entropy rather than on the quantum relative entropy itself. The use of a quantum entropy for the purpose of inference has a large history: Jaynes \cite{Jaynes1 1957,Jaynes book} invented the notion of the quantum maximum entropy method \cite{Jaynes2 1957}, while it was perpetuated by \cite{Balian1,Balian2,Balian3, Balian4, Balian5, Balian6,Partovi1, Partovi2} and many others. However, we find the quantum \emph{relative} entropy to be the suitable entropy for updating density matrices, rather than the von Neumann. The relevant results of their papers may be found using our quantum relative entropy with a suitable uniform prior density matrix. 

It should be noted that because the relative entropies were reached by design, they may be interpret as such, ``the relative entropies are tools for updating", which means we no longer need to attach an interpretation \emph{ex post facto} -- as a measure of disorder or amount of missing information. In this sense, the relative entropies were built for the purpose of saturating their own interpretation \cite{Skilling1,book}.

The remainder of the paper is organized as follows: First we will discuss some universally applicable principles of inference and motivate the design of an entropy function able to rank probability distributions. This entropy function will be designed such it is consistent with inference by applying a few reasonable design criteria, which are guided by the aforementioned principles of inference. Using the same principles of inference and design criteria, we find the form of the quantum relative entropy suitable for inference. We end with concluding remarks. 

Solutions for $\hat{\rho}$ by maximizing the quantum relative entropy give insight into the Quantum Bayes' Rule in the sense of \cite{Korotkov1, Korotkov2, Jordan, quantum bayes}. This, and a few other applications of the quantum maximum entropy method, will be discussed in a future article. 
%

\section{The Design of Entropic Inference}

Inference is the appropriate updating of probability distributions when new information is received. Bayes' rule and Jeffrey's rule are both equipped to handle information in the form of data; however, the updating of a probability distribution due to the knowledge of an expectation value was realized by Jaynes \cite{Jaynes1 1957, Jaynes book, Jaynes2 1957} through the method of maximum entropy. The two methods for inference were thought to be devoid of one another until the work of \cite{Giffin}, which showed Bayes' and Jeffrey's Rule to be consistent with the method of maximum entropy when the expectation values were in the form of data \cite{Giffin}. In the spirit of the derivation we will carry-on as if the maximum entropy method were not known and show how it may be derived as an application of inference. 

Given a probability distribution $\varphi(x)$ over a general set of propositions $x\in X$, it is self evident that if new information is learned, we are entitled to assign a new probability distribution $\rho(x)$ that somehow reflects this new information while also respecting our prior probability distribution $\varphi(x)$. The main question we must address is: ``Given some information, to what posterior probability distribution $\rho(x)$ should we update our prior probability distribution $\varphi(x)$ to?", that is,
\begin{eqnarray}
\varphi(x)\stackrel{*}{\longrightarrow}\rho(x)?\nonumber
\end{eqnarray}
This specifies the problem of inductive inference.  Since ``information" has many colloquial, yet potentially conflicting, definitions, we remove potential confusion by defining {\bf information} operationally $(*)$ as the \emph{rationale} that causes a probability distribution to change (inspired by and adapted from \cite{book}). Directly from \cite{book}:\\\\
``Our goal is to design a method that allows a systematic search for the preferred
posterior distribution. The central idea, first proposed in \cite{Skilling1}
is disarmingly simple: to select the posterior first rank all candidate distributions
in increasing \emph{order of preference} and then pick the distribution that ranks
the highest. Irrespective of what it is that makes one distribution preferable over
another (we will get to that soon enough) it is clear that any ranking according
to preference must be transitive: if distribution $\rho_1$ is preferred over distribution
$\rho_2$, and $\rho_2$ is preferred over $\rho_3$, then $\rho_1$ is preferred over $\rho_3$. Such transitive
rankings are implemented by assigning to each $\rho(x)$ a real number $S[\rho]$, which
is called the entropy of $\rho$, in such a way that if $\rho_1$ is preferred over $\rho_2$, then
$S[\rho_1] > S[\rho_2]$. The selected distribution (one or possibly many, for there may
be several equally preferred distributions) is that which maximizes the entropy
functional." \\

Because we wish to update from prior distributions $\varphi$ to posterior distributions $\rho$ by ranking, the entropy functional $S[\rho,\varphi]$, is a real function of both $\varphi$ and $\rho$.  In the absence of new information, there is no available \emph{rationale} to prefer any $\rho$ to the original $\varphi$, and thereby the relative entropy should be designed such that the selected posterior is equal to the prior $\varphi$ (in the absence of new information). The prior information encoded in $\varphi(x)$ is valuable and we should not change it unless we are informed otherwise. Due to our definition of information, and our desire for objectivity, we state the predominate guiding principle for inductive inference:

\paragraph{The Principle of Minimal Updating (PMU):}\emph{A probability distribution should only be updated to the extent required by the new information.} \\

This simple statement provides the foundation for inference \cite{book}. If the updating of probability distributions is to be done objectively, then possibilities should not be needlessly ruled out or suppressed. Being informationally stingy, that we should only update probability distributions when the information requires it, pushes inductive inference toward objectivity. Thus using the PMU helps formulate a pragmatic (and objective) procedure for making inferences using (informationally) subjective probability distributions \cite{Pragmatic}. 

This method of inference is only as universal and general as its ability to apply \emph{equally well} to \emph{any} specific inference problem. The notion of ``specificity" is the notion of statistical independence; a special case is only special in that it is separable from other special cases. The notion that systems may be ``sufficiently independent" plays a central and deep-seated role in science and the idea that some things can be neglected and that not everything matters, is implemented by imposing criteria that tells us how to handle independent systems \cite{book}. Ironically, the universally \emph{shared} property by all specific inference problems is their ability to be \emph{independent} of one another. Thus, a universal inference scheme based on the PMU permits,

\paragraph{Properties of Independence (PI):} \emph{Subdomain Independence: When information is received about one set of propositions, it should not effect or change the state of knowledge (probability distribution) of the other propositions (else information was also received about them too);} 
\begin{eqnarray}
\mbox{\emph{And,}}\nonumber
\end{eqnarray}

\emph{Subsystem Independence: When two systems are a-priori believed to be
independent and we only receive information about one, then the state of knowledge of the other system remains unchanged.}\\

The PI's are special cases of the PMU that ultimately take the form of \emph{design criteria} in the design derivation.  The process of constraining the form of $S[\rho,\varphi]$ by imposing design criteria may be viewed as the process of \emph{eliminative induction}, and after sufficient constraining, a single form for the entropy remains. Thus, the justification behind the surviving entropy is not that it leads to demonstrably correct inferences, but rather, that all other candidate entropies demonstrably fail to perform as desired \cite{book}. Rather than the \emph{design criteria} instructing one how to update, they instruct in what instances one should \emph{not} update. That is, rather than justifying one way to skin a cat over another, we tell you when \emph{not} to skin it, which is operationally unique -- namely you don't do it -- luckily enough for the cat.

\subsection{The Design Criteria and the Standard Relative Entropy}
The following \emph{design criteria} (DC), guided by the PMU, are imposed and formulate the standard relative entropy as a tool for inference. The form of this presentation is inspired by \cite{book}.
\begin{description}
\item[\textbf{DC1: Subdomain Independence}] 
\end{description}

\noindent We keep the DC1 from \cite{book} and review it below. DC1 imposes the first instance of when one should not update -- the Subdomain PI.  Suppose the information to be processed does \emph{not }refer to a
particular subdomain $\mathcal{D}$ of the space $\mathcal{X}$ of $x$'s. In
the absence of new information about $\mathcal{D}$ the PMU insists we do
not change our minds about probabilities that are conditional on $\mathcal{D}
$. Thus, we design the inference method so that $\varphi(x|\mathcal{D})$, the
prior probability of $x$ conditional on $x\in\mathcal{D}$, is not updated and therefore the selected conditional posterior is,
\begin{equation}
P(x|\mathcal{D})=\varphi(x|\mathcal{D}).  \label{DC1a}
\end{equation}
(The notation will be as follows: we denote priors by $\varphi$, candidate
posteriors by lower case $\rho$, and the selected posterior by upper case $P$.)  We emphasize the point is not that we make the unwarranted assumption that
keeping $\varphi(x|\mathcal{D})$ unchanged is guaranteed to lead to correct
inferences. It need not; induction is risky. The point is, rather, that in
the absence of any evidence to the contrary there is no reason to change our
minds and the prior information takes priority.

 \begin{description}
\item[\textbf{DC1 Implementation:}] 
\end{description}
Consider the set of microstates $x_i\in \mathcal{X}$ belonging to either of two non-overlapping domains $\mathcal{D}$ or its compliment $\mathcal{D}'$, such that $\mathcal{X}=\mathcal{D}\cup \mathcal{D}'$ and $\emptyset=\mathcal{D}\cap \mathcal{D}'$. For convenience let $\rho(x_i)=\rho_i$.  Consider the following constraints:
\begin{eqnarray}
\rho(\mathcal{D})=\sum_{i\in \mathcal{D}}\rho_i\quad\mbox{and}\quad \rho(\mathcal{D}')=\sum_{i\in \mathcal{D}'}\rho_i,\label{DC1}
\end{eqnarray}
such that $\rho(\mathcal{D})+\rho(\mathcal{D}')=1$, and the following ``local"  constraints to $\mathcal{D}$ and $\mathcal{D}'$ respectively are,
\begin{eqnarray}
\expt{A}=\sum_{i\in \mathcal{D}}\rho_iA_i\quad\mbox{and}\quad \expt{A'}=\sum_{i\in \mathcal{D}'}\rho_iA'_i.\label{DC2}
\end{eqnarray}
As we are searching for the candidate distribution which maximizes $S$ while obeying (\ref{DC1}) and (\ref{DC2}), we maximize the entropy $S\equiv S[\rho,\varphi]$ with respect to these expectation value constraints using the Lagrange multiplier method,
\begin{eqnarray}
0=\delta\Big(S-\lambda[\rho(\mathcal{D})-\sum_{i\in \mathcal{D}}\rho_i]-\mu[\expt{A}-\sum_{i\in \mathcal{D}}\rho_iA_i]\nonumber\\
-\lambda'[\rho(\mathcal{D}')-\sum_{i\in \mathcal{D}'}\rho_i]-\mu'[\expt{A'}-\sum_{i\in \mathcal{D}'}\rho_iA_i]\Big),\nonumber
\end{eqnarray}
 and thus, the entropy is maximized when the following differential relationships hold:
\begin{eqnarray}
\frac{\delta S}{\delta \rho_i}&=&\lambda+\mu A_i\quad\mbox{$\forall\, i\in \mathcal{D}$},\label{c3}\\
\frac{\delta S}{\delta \rho_i}&=&\lambda'+\mu' A'_i\quad\mbox{$\forall\, i\in \mathcal{D}'$}.\label{c4}
\end{eqnarray}
Equations (\ref{DC1})-(\ref{c4}), are $n+4$ equations we must solve to find the four Lagrange multipliers $\{\lambda,\lambda',\mu,\mu'\}$ and the $n$ probability values $\{\rho_i\}$.

If the subdomain constraint DC1 is imposed in the most restrictive case, then it will hold in general. The most restrictive case requires splitting $\mathcal{X}$ into a set of $\{\mathcal{D}_i\}$ domains such that each $\mathcal{D}_i$ singularly includes one microstate $x_i$. This gives,
\begin{eqnarray}
\frac{\delta S}{\delta \rho_i}=\lambda_i+\mu_iA_i\quad\mbox{in each $\mathcal{D}_i$}.\label{dc1 7}
\end{eqnarray}
Because the entropy $S=S[\rho_1,\rho_2,...;\varphi_1,\varphi_2,...]$ is a function over the probability of each microstate's posterior and prior distribution, its variational derivative is also a function of said probabilities in general,
\begin{eqnarray}
\frac{\delta S}{\delta \rho_i}\equiv\phi_i(\rho_1,\rho_2,...;\varphi_1,\varphi_2,...)=\lambda_i+\mu_iA_i\quad\mbox{for each $(i,\mathcal{D}_i)$}.
\end{eqnarray}
DC1 is imposed by constraining the form of $\phi_i(\rho_1,\rho_2,...;\varphi_1,\varphi_2,...)=\phi_i(\rho_i;\varphi_1,\varphi_2,...)$ to ensures that changes in $A_i\rightarrow A_i+\delta A_i$ have no influence over the value of $\rho_j$ in domain $\mathcal{D}_j$, through $\phi_i$, for $i\neq j$. If there is no new information about propositions in $\mathcal{D}_j$, its distribution should remain equal to $\varphi_j$ by the PMU. We further restrict $\phi_i$ such that an arbitrary variation of $\varphi_j\rightarrow \varphi_j+\delta \varphi_j$ (a change in the prior state of knowledge of the microstate $j$) has no effect on $\rho_i$ for $i\neq j$ and therefore DC1 imposes $\phi_i=\phi_i(\rho_i,\varphi_i)$, as is guided by the PMU. At this point it is easy to generalize the analysis to continuous microstates such that the indices become continuous $i\rightarrow x$, sums become integrals, and discrete probabilities become probability densities $\rho_i\rightarrow \rho(x)$.

 \begin{description}
\item[\noindent \textbf{Remark:}] 
\end{description}
We are designing the entropy for the purpose of ranking posterior probability distributions (for the purpose of inference); however, the highest ranked distribution is found by setting the variational derivative of $S[\rho,\varphi]$ equal to the variations of the expectation value constraints by the Lagrange multiplier method,
\begin{eqnarray}
\frac{\delta S}{\delta \rho(x)}=\lambda+\sum_i\mu_i A_i(x).
\end{eqnarray}
Therefore, the real quantity of interest is $\frac{\delta S}{\delta \rho(x)}$ rather than the specific form of $S[\rho,\varphi]$. \emph{All} forms of $S[\rho,\varphi]$ that give the correct form of $\frac{\delta S}{\delta \rho(x)}$ are \emph{equally valid} for the purpose of inference. Thus, every design criteria may be made on the variational derivative of the entropy rather than the entropy itself, which we do. When maximizing the entropy, for convenience, we will let,
\begin{eqnarray}
\frac{\delta S}{\delta \rho(x)}\equiv \phi_x(\rho(x),\varphi(x)),
\end{eqnarray}
and further use the shorthand $\phi_x(\rho,\varphi)\equiv\phi_x(\rho(x),\varphi(x))$, in all cases. 

\begin{description}
\item[\textbf{DC1':}] \emph{In the absence of new information, our new state of knowledge $\rho(x)$ is equal to the old state of knowledge $\varphi(x)$.}
\end{description}

This is a special case of DC1, and is implemented differently than in \cite{book}. The PMU is in principle a statement about informational honestly -- that is, one should not ``jump to conclusions" in light of new information and in the absence of new information, one should not change their state of knowledge. If no new information is given, the prior probability distribution $\varphi(x)$ does not change, that is, the posterior probability distribution $\rho(x)=\varphi(x)$ is equal to the prior probability. If we maximizing the entropy without applying constraints,
\begin{eqnarray}
\frac{\delta S}{\delta \rho(x)}=0,
\end{eqnarray}
then DC1' imposes the following condition:
\begin{eqnarray}
\frac{\delta S}{\delta \rho(x)}=\phi_x(\rho,\varphi)=\phi_x(\varphi,\varphi)=0,\label{18}
\end{eqnarray}
for all $x$ in this case. This special case of the DC1 and the PMU turns out to be incredibly constraining as we will see over the course of DC2.

\noindent \textbf{Comment:}

From \cite{book}. If the variable $x$ is continuous, DC1 requires that when information refers to points infinitely close but just outside the
domain $\mathcal{D}$, that it will have no influence on probabilities conditional on $%
\mathcal{D}$. This may seem surprising as it may lead to updated probability
distributions that are discontinuous. Is this a problem? No.

In certain situations (\emph{e.g.}, physics) we might have explicit reasons
to believe that conditions of continuity or differentiability should be
imposed and this information might be given to us in a variety of ways. The
crucial point, however -- and this is a point that we keep and will keep
reiterating -- is that unless such information is explicitly given
we should not assume it. If the new information leads to discontinuities, so
be it.\\

%

\begin{description}
\item[\textbf{DC2: Subsystem Independence}] 
\end{description}

DC2 imposes the second instance of when one should not update -- the Subsystem PI. We emphasize that \emph{DC2 is not a consistency requirement}. The
argument we deploy is \emph{not} that both the prior \emph{and} the new
information tells us the systems are independent, in which case consistency
requires that it should not matter whether the systems are treated jointly
or separately. Rather, DC2 refers to a situation where the new information does not
say whether the systems are independent or not, but information is given about each subsystem. The updating is
being \emph{designed} so that the independence reflected in the prior is
maintained in the posterior by default via the PMU and the second clause of the PI's. \cite{book}

The point is not that when we have no evidence for
correlations we draw the firm conclusion that the systems must necessarily
be independent. They could indeed have turned out to be correlated and then
our inferences would be wrong. Again, induction involves risk. The point is rather
that if the joint prior reflected independence and the new evidence is
silent on the matter of correlations, then the prior takes precedence. As before, in this case subdomain independence, the probability distribution should not be
updated unless the information requires it. \cite{book}

\begin{description}
\item[\textbf{DC2 Implementation:}]
\end{description}
Consider a composite system, $x=(x_{1},x_{2})\in \mathcal{X}=\mathcal{X}%
_{1}\times \mathcal{X}_{2}$. Assume that all prior evidence led us to
believe the subsystems are independent. This belief is reflected in the prior
distribution: if the individual system priors are $\varphi_{1}(x_{1})$ and $%
\varphi_{2}(x_{2})$, then the prior for the whole system is their product $%
\varphi_{1}(x_{1})\varphi_{2}(x_{2})$. Further suppose that new information is acquired
such that $\varphi_{1}(x_{1})$ would by itself be updated to $P_{1}(x_{1})$ and
that $\varphi_{2}(x_{2})$ would be itself be updated to $P_{2}(x_{2})$. By design, the implementation of DC2
constrains the entropy functional such that in this case, the joint product prior $%
\varphi_{1}(x_{1})\varphi_{2}(x_{2}) $ updates to the selected product posterior $P_{1}(x_{1})P_{2}(x_{2})$. \cite{book}

The argument below is considerably simplified if we expand the space of
probabilities to include distributions that are not necessarily normalized. This does
not represent any limitation because a normalization constraint may always be applied. We consider a few special cases below: \\

\noindent \textbf{Case 1:} We receive the extremely constraining information
that the posterior distribution for system $1$ is completely specified to be 
$P_{1}(x_{1})$ while we receive no information at all about system $2$. We
treat the two systems jointly. Maximize the joint entropy $S[\rho(x_1,x_2),\varphi(x_1)\varphi(x_2)]$
subject to the following constraints on the $\rho(x_{1},x_{2})\,$, 
\begin{equation}
\int dx_{2}\,\rho(x_{1},x_{2})=P_{1}(x_{1})~.
\end{equation}%
Notice that the probability of each $x_{1}\in\mathcal{X}_1$ within $\rho(x_1,x_2)$ is being constrained to $P_{1}(x_{1})$ in the marginal. We therefore need a one Lagrange multiplier $\lambda _{1}(x_{1})$ for each $x_1\in\mathcal{X}_1$ to tie each value of $\int dx_2 \,\rho(x_1,x_2)$ to $P_1(x_1)$. Maximizing the entropy with respect to this constraint is,
\begin{equation}
\delta \left[ S-\int dx_{1}\lambda _{1}(x_{1})\left( \int
dx_{2}\,\rho(x_{1},x_{2})-P_{1}(x_{1})\right) \right] =0\,,
\end{equation}%
which requires that
\begin{equation}
\lambda _{1}(x_{1})=\phi _{x_{1}x_{2}}\left( \rho(x_{1},x_{2}),\varphi_{1}(x_{1})\varphi_{2}(x_{2})\right)\,,
\end{equation}%
for arbitrary variations of $\rho(x_1,x_2)$. By design, DC2 is implemented by requiring $\varphi_{1}\varphi_{2}\rightarrow P_{1}\varphi_{2}$ in this case, therefore,
\begin{equation}
\lambda _{1}(x_{1})=\phi _{x_{1}x_{2}}\left(
P_{1}(x_{1})\varphi_{2}(x_{2}),\varphi_{1}(x_{1})\varphi_{2}(x_{2})\right) ~.
\label{lambda1 a}
\end{equation}%
This equation must hold for all choices of $%
x_{2}$ and all choices of the prior $\varphi_{2}(x_{2})$ as $\lambda _{1}(x_{1})$ is independent of $x_{2}$. Suppose we had chosen a different prior $\varphi_{2}^{\prime }(x_{2})=\varphi_{2}(x_{2})+\delta \varphi_{2}(x_{2})$ that
disagrees with $\varphi_{2}(x_{2})$. For all $x_2$ and $\delta \varphi_{2}(x_{2})$, the multiplier $\lambda _{1}(x_{1})$ remains unchanged as it constrains the independent $\rho(x_1)\rightarrow P_1(x_1)$. This means that any dependence that the right hand side might potentially have had on $x_{2}$ and on the
prior $\varphi_{2}(x_{2})$ \emph{must cancel out}. This means that
\begin{equation}
\phi _{x_{1}x_{2}}\left(P_{1}(x_{1})\varphi_{2}(x_{2}),\varphi_{1}(x_{1})\varphi_{2}(x_{2})\right)=f_{x_1}(P_{1}(x_{1}),\varphi_{1}(x_{1})).
\end{equation}
Since $\varphi_{2}$ is arbitrary in $f$
suppose further that we choose a constant prior set equal to one, $\varphi_{2}(x_{2})=1$, therefore
\begin{equation}
f_{x_1}(P_{1}(x_{1}),\varphi_{1}(x_{1}))=\phi _{x_{1}x_2}\left(
P_{1}(x_{1})*1,\varphi_{1}(x_{1})*1\right)=\phi _{x_{1}}\left(
P_{1}(x_{1}),\varphi_{1}(x_{1})\right)
\end{equation}
 in general. This gives, 
\begin{equation}
\lambda _{1}(x_{1})=\phi _{x_{1}}\left(
P_{1}(x_{1}),\varphi_{1}(x_{1})\right) .  \label{lambda1 b}
\end{equation}%
The left hand side does not depend on $x_{2}$, and therefore neither does
the right hand side. An argument exchanging systems $1$ and $2$ gives a similar result.

\noindent \textbf{Case 1 - Conclusion:} When the system 2 is not updated the dependence on $\varphi_{2}$ and $x_2$
drops out,     
\begin{equation}
\phi_{x_1x_2} \left( P_{1}(x_1)\varphi_{2}(x_2),\varphi_{1}(x_1)\varphi_{2}(x_2)\right) =\phi_{x_1} \left( P_{1}(x_1),\varphi_{1}(x_1)\right) .
\label{lambda1 c}
\end{equation}
and vice-versa when system 1 is not updated,
\begin{equation}
\phi_{x_1x_2} \left( \varphi_{1}(x_1)P_{2}(x_2),\varphi_{1}(x_1)\varphi_{2}(x_2)\right) =\phi_{x_2} \left( P_{2}(x_2),\varphi_{2}(x_2)\right) .\label{lambda1 d}
\end{equation}
As we seek the general functional form of $\phi_{x_1x_2}$, and because the $x_2$ dependence drops out of (\ref{lambda1 c}) and the $x_1$ dependence drops out of (\ref{lambda1 d}) for arbitrary $\varphi_1,\varphi_2$ and $\varphi_{12}=\varphi_1\varphi_2$, the explicit coordinate dependence in $\phi$ consequently drops out of both such that,
\begin{eqnarray}
\phi_{x_1x_2}\rightarrow \phi,
\end{eqnarray}
as $\phi=\phi(\rho(x),\varphi(x))$ must only depend on coordinates through the probability distributions themselves.  (As a double check, explicit coordinate dependence was included in the following computations but inevitably dropped out due to the form the functional equations and DC1'. By the argument above, and for simplicity, we drop the explicit coordinate dependence  in $\phi$ here.)\\

\noindent \textbf{Case 2:} Now consider a different special case in which
the marginal posterior distributions for systems $1$ and $2$ are both
completely specified to be $P_{1}(x_{1})$ and $P_{2}(x_{2})$ respectively.
Maximize the joint entropy $S[\rho(x_1,x_2),\varphi(x_1)\varphi(x_2)]$
subject to the following constraints on the $\rho(x_{1},x_{2})\,$, 
\begin{equation}
\int dx_{2}\,\rho(x_{1},x_{2})=P_{1}(x_{1})\quad \mbox{and}\quad \int
dx_{1}\,\rho(x_{1},x_{2})=P_{2}(x_{2})~.
\end{equation}%
Again, this is one constraint for each value of $x_{1}$ and one constraint for each value of $x_{2}$, which therefore require the separate multipliers $\mu _{1}(x_{1})$ and $\mu
_{2}(x_{2})$. Maximizing $S$ with respect to these constraints is then, 
\begin{eqnarray}
0 &=&\delta \left[ S-\int dx_{1}\mu _{1}(x_{1})\left( \int
dx_{2}\,\rho(x_{1},x_{2})-P_{1}(x_{1})\right) \right.   \nonumber \\
&&-\left. \int dx_{2}\mu _{2}(x_{2})\left( \int
dx_{1}\,\rho(x_{1},x_{2})-P_{2}(x_{2})\right) \right] \,,
\end{eqnarray}%
leading to

\begin{equation}
\mu
_{1}(x_{1})+\mu _{2}(x_{2})=\phi \left( \rho(x_{1},x_{2}),\varphi_{1}(x_{1})\varphi_{2}(x_{2})\right).
\end{equation}%
The updating is being designed so that $\varphi_{1}\varphi_{2}\rightarrow P_{1}P_{2}$, as the independent subsystems are being updated based on expectation values which are silent about correlations. DC2 thus imposes,
\begin{equation}
\mu _{1}(x_{1})+\mu _{2}(x_{2})=\phi \left(
P_{1}(x_{1})P_{2}(x_{2}),\varphi_{1}(x_{1})\varphi_{2}(x_{2})\right) .  \label{mu12 a}
\end{equation}%
Write (\ref{mu12 a}) as,
\begin{equation}
\mu _{1}(x_{1})=\phi \left(
P_{1}(x_{1})P_{2}(x_{2}),\varphi_{1}(x_{1})\varphi_{2}(x_{2})\right)-\mu _{2}(x_{2}) .  \label{mu12 b}
\end{equation}%
The left hand side is independent of $x_{2}$ so we can perform a trick
similar to that we used before. Suppose we had chosen a different \emph{%
constraint} $P_{2}^{\prime }(x_{2})$ that differs from $P_{2}(x_{2})$ and a new prior $\varphi'_2(x_{2})$ that differs from $\varphi_2(x_{2})$ except at the value $\bar{x}_2$. At the value $\bar{x}_2$,the multiplier $\mu
_{1}(x_{1})$ remains unchanged for all $P_{2}^{\prime }(x_{2})$, $\varphi'_2(x_{2})$, and thus $x_2$. This means that any dependence that the right hand side might potentially have had on $x_{2}$ and on the choice of $P_{2}(x_{2})$, $\varphi'_2(x_{2})$ must cancel out leaving $\mu _{1}(x_{1})$ unchanged. That is, the Lagrange multiplier $\mu(x_2)$ ``pushes out" these dependences such that 
\begin{equation}
\phi \left(
P_{1}(x_{1})P_{2}(x_{2}),\varphi_{1}(x_{1})\varphi_{2}(x_{2})\right)-\mu _{2}(x_{2})=g(P_{1}(x_{1}),\varphi_{1}(x_{1})).
\end{equation}
Because $g(P_{1}(x_{1}),\varphi_{1}(x_{1}))$ is independent of arbitrary variations of $P_{2}(x_{2})$ and $\varphi_{2}(x_{2})$ on the LHS above -- it is satisfied equally well for all choices.  The form of $g=\phi(P_{1}(x_{1}),q_{1}(x_{1}))$ is apparent if $P_{2}(x_{2})=\varphi_{2}(x_{2})=1$ as $\mu_2(x_2)=0$ similar to Case 1 as well as DC1'. Therefore, the Lagrange multiplier is 
\begin{equation}
\mu _{1}(x_{1})=\phi\left(
P_{1}(x_{1}),\varphi_{1}(x_{1})\right) .
\end{equation}%
A similar analysis can be carried out for $\mu _{2}(x_{2})$ leads to  
\begin{equation}
\mu _{2}(x_{2})=\phi \left( P_{2}(x_{2}),\varphi_{2}(x_{2})\right) .
\end{equation}
\noindent \textbf{Case 2 - Conclusion:}
Substituting back into (\ref{mu12 a}) gives us a functional equation for $%
\phi \,$,     
\begin{equation}
\phi \left( P_{1}P_{2},\varphi_{1}\varphi_{2}\right) =\phi \left( P_{1},\varphi_{1}\right)
+\phi \left( P_{2},\varphi_{2}\right) .\label{funcequationprob}
\end{equation}%
The general solution for this functional equation is derived in the Appendix, section \ref{appendix1}, and is
\begin{eqnarray}
\phi(\rho,\varphi)=a_1\ln(\rho(x))+a_2\ln(\varphi(x))
\end{eqnarray}
where $a_1,a_2$ are constants. The constants are fixed by using DC1'. Letting $\rho_1(x_1)=\varphi_1(x_1)=\varphi_1$ gives $\phi(\varphi,\varphi)=0$ by DC1', and therefore,
\begin{eqnarray}
\phi(\varphi,\varphi)=(a_1+a_2)\ln(\varphi)=0,
\end{eqnarray}
so we are forced to conclude $a_1=-a_2$ for arbitrary $\varphi$.
Letting $a_1\equiv A=-|A|$ such that we are really maximizing the entropy (although this is purely aesthetic) gives the general form of $\phi$ to be,
\begin{eqnarray}
\phi(\rho,\varphi)=-|A|\ln\Big(\frac{\rho(x)}{\varphi(x)}\Big).
\end{eqnarray}
As long as $A\neq0$, the value of $A$ is arbitrary as it always can be absorbed into the Lagrange multipliers. The general form of the entropy designed for the purpose of inference of $\rho$ is found by integrating $\phi$, and therefore,
\begin{eqnarray}
S(\rho(x),\varphi(x))=-|A|\int dx\, (\rho(x)\ln\Big(\frac{\rho(x)}{\varphi(x)}\Big)-\rho(x))+C[\varphi].\label{theentropy}
\end{eqnarray}
The constant in $\rho$, $C[\varphi]$, will always drop out when varying $\rho$.
The apparent extra term ($|A|\int\rho(x)dx$) from integration cannot be dropped while simultaneously satisfying DC1', which requires $\rho(x)=\varphi(x)$ in the absence of constraints or when there is no change to one's information. In previous versions where the integration term ($|A|\int\rho(x)dx$) is dropped, one obtains solutions like $\rho(x)=e^{-1}\varphi(x)$ (independent of whether $\varphi(x)$ was previously normalized or not) in the absence of new information. Obviously this factor can be taken care of by normalization, and in this way both forms of the entropy are equally valid; however, this form of the entropy better adheres to the PMU through DC1'. Given that we may regularly impose normalization, we may drop the extra $\int \rho(x) dx$ term and $C[\varphi]$. For convenience then, (\ref{theentropy}) becomes 
\begin{eqnarray}
S(\rho(x),\varphi(x))\rightarrow S^*(\rho(x),\varphi(x))=-|A|\int dx\, \rho(x)\ln\Big(\frac{\rho(x)}{\varphi(x)}\Big),
\end{eqnarray}
which is a special case when the normalization constraint is being applied. Given normalization is applied, the same selected posterior $\rho(x)$ maximizes both $S(\rho(x),\varphi(x))$ and $S^*(\rho(x),\varphi(x))$, and the star notation may be dropped.

\paragraph{Remarks:}
It can be seen that the relative entropy is invariant under coordinate transformations. This implies that a system of coordinates carry no information and it is the ``character" of the probability distributions that are being ranked against one another rather than the specific set of propositions or microstates they describe.

The general solution to the maximum entropy procedure with respect to $N$ linear constraints in $\rho$, $\expt{A_i(x)}$, and normalization gives a canonical-like selected posterior probability distribution,
\begin{eqnarray}
\rho(x)=\varphi(x)\exp\Big(\sum_i\alpha_iA_i(x)\Big).\label{canonical}
\end{eqnarray}
 The positive constant $|A|$ may always be absorbed into the Lagrange multipliers so we may let it equal unity without loss of generality. DC1' is fully realized when we maximize with respect to a constraint on $\rho(x)$ that is already held by $\varphi(x)$, such as $\expt{x^2}=\int x^2\rho(x)$ which happens to have the same value as $\int x^2\varphi(x)$, then its Lagrange multiplier is forcibly zero $\alpha_{1}=0$ (as can be seen in (\ref{canonical}) using (\ref{theentropy})), in agreement with Jaynes. This gives the expected result $\rho(x)=\varphi(x)$ as there is no new information. 
Our design has arrived at a refined maximum entropy method \cite{Jaynes1 1957} as a universal probability updating procedure \cite{Giffin}. 

%

\section{The Design of the Quantum Relative Entropy}
 Last section we assumed that the universe of discourse (the set of relevant propositions or microstates) $\mathcal{X}=\mathcal{A}\times\mathcal{B}\times...$ was known. In quantum physics things are a bit more ambiguous because many probability distributions, or many experiments, can be associated to a given density matrix. In this sense it helpful to think of density matrices as ``placeholders" for probability distributions rather than a probability distributions themselves. As any probability distribution from a given density matrix, $\rho(\cdot)=\mbox{Tr}(\ket{\cdot}\bra{\cdot}\hat{\rho})$, may be ranked using the standard relative entropy, it is unclear why we would chose one universe of discourse over another. In lieu of this, such that one universe of discourse is not given preferential treatment, we consider ranking entire density matrices against one another. Probability distributions of interest may be found from the selected posterior density matrix. This moves our universe of discourse from sets of propositions $\mathcal{X}\rightarrow\mathcal{H}$ to Hilbert space(s).

When the objects of study are quantum systems, we desire an objective procedure to update from a prior density matrix $\hat{\varphi}$ to a posterior density matrix $\hat{\rho}$. We will apply the same intuition for ranking probability distributions (Section 2) and implement the PMU, PI, and design criteria to the ranking of density matrices. We therefore find the quantum relative entropy $S(\hat{\rho},\hat{\varphi})$ to be designed for the purpose of inferentially updating density matrices.

\subsection{Designing the Quantum Relative Entropy}
In this section we design the quantum relative entropy using the same inferentially guided \emph{design criteria} as were used in the standard relative entropy. 
\begin{description}
\item[\textbf{DC1: Subdomain Independence}] 
\end{description}



The goal is to design a function $S(\hat{\rho},\hat{\varphi})$ which is able to rank density matrices. This insists that $S(\hat{\rho},\hat{\varphi})$ be a real scalar valued function of the posterior $\hat{\rho}$, and prior $\hat{\varphi}$ density matrices, which we will call the quantum relative entropy or simply the entropy. An arbitrary variation of the entropy with respect to $\hat{\rho}$ is,
\begin{eqnarray}
\delta\,S(\hat{\rho},\hat{\varphi})=\sum_{ij}\frac{\delta S(\hat{\rho},\hat{\varphi})}{\delta \rho_{ij}}\delta\rho_{ij}=\sum_{ij}\Big(\frac{\delta S(\hat{\rho},\hat{\varphi})}{\delta \hat{\rho}}\Big)_{ij}\delta(\hat{\rho})_{ij}=\sum_{ij}\Big(\frac{\delta S(\hat{\rho},\hat{\varphi})}{\delta \hat{\rho}^T}\Big)_{ji}\delta(\hat{\rho})_{ij}=\mbox{Tr}\Big(\frac{\delta S(\hat{\rho},\hat{\varphi})}{\delta \hat{\rho}^T}\delta\hat{\rho}\Big).
\end{eqnarray}
We wish to maximize this entropy with respect to expectation value constraints, such as, $\expt{A}=\mbox{Tr}(\hat{A}\hat{\rho})$ on $\hat{\rho}$. Using the Lagrange multiplier method to maximize the entropy with respect to $\expt{A}$ and normalization, is setting the variation equal to zero,

\begin{eqnarray}
\delta\Big(S(\hat{\rho},\hat{\varphi})-\lambda[\mbox{Tr}(\hat{\rho})-1]-\alpha[\mbox{Tr}(\hat{A}\hat{\rho})-\expt{A}]\Big)=0,
\end{eqnarray}
 where $\lambda$ and $\alpha$ are the Lagrange multipliers for the respective constraints. Because $S(\hat{\rho},\hat{\varphi})$ is a real number, we inevitably require $\delta S$ to be real, but without imposing this directly, we find that requiring $\delta S$ to be real requires $\hat{\rho},\hat{A}$ to be Hermitian. At this point, it is simpler to allow for arbitrary variations of $\hat{\rho}$ such that,
\begin{eqnarray}
\mbox{Tr}\Big(\Big(\frac{\delta S(\hat{\rho},\hat{\varphi})}{\delta \hat{\rho}^T}-\lambda\hat{1}-\alpha\hat{A}\Big)\delta\hat{\rho}\Big)=0.
\end{eqnarray}
For these arbitrary variations, the variational derivative of $S$ must satisfy,
\begin{eqnarray}
\frac{\delta S(\hat{\rho},\hat{\varphi})}{\delta \hat{\rho}^T}=\lambda\hat{1}+\alpha\hat{A},\label{52}
\end{eqnarray}
at the maximum. As in the remark earlier, \emph{all} forms of $S$ which give the correct form of $\frac{\delta S(\hat{\rho},\hat{\varphi})}{\delta \hat{\rho}^T}$ under variation are \emph{equally valid} for the purpose of inference. For notational convenience we let,
\begin{eqnarray}
\frac{\delta S(\hat{\rho},\hat{\varphi})}{\delta \hat{\rho}^T}\equiv\phi(\hat{\rho},\hat{\varphi}),
\end{eqnarray}
which is a matrix valued function of the posterior and prior density matrices. The form of $\phi(\hat{\rho},\hat{\varphi})$ is already "local" in $\hat{\rho}$, so we don't need to constrain it further as we did in the original DC1. 

\begin{description}
\item[\textbf{DC1':}] \emph{In the absence of new information, the new state $\hat{\rho}$ is equal to the old state $\hat{\varphi}$.}
\end{description}

 Applied to the ranking of density matrices, in the absence of new information, the density matrix $\hat{\varphi}$ should not change, that is, the posterior density matrix $\hat{\rho}=\hat{\varphi}$ is equal to the prior density matrix. Maximizing the entropy without applying any constraints gives,
\begin{eqnarray}
\frac{\delta S(\hat{\rho},\hat{\varphi})}{\delta \hat{\rho}^T}=\hat{0},
\end{eqnarray}
and therefore DC1' imposes the following condition in this case,
\begin{eqnarray}
\frac{\delta S(\hat{\rho},\hat{\varphi})}{\delta \hat{\rho}^T}=\phi(\hat{\rho},\hat{\varphi})=\phi(\hat{\varphi},\hat{\varphi})=\hat{0}.
\end{eqnarray}
As in the original DC1', if $\hat{\varphi}$ is known to obey some expectation value constraint $\expt{\hat{A}}$, then if one goes out of their way to constrain $\hat{\rho}$ to that expectation value with nothing else, it follows from the PMU that $\hat{\rho}=\hat{\varphi}$, as no information has been gained. This is not imposed directly, but can be verified later.

\begin{description}
\item[\textbf{DC2: Subsystem Independence}]
\end{description}

The discussion of DC2 is the same as the standard relative entropy DC2 -- it is not a consistency requirement, and the updating is
 \emph{designed} so that the independence reflected in the prior is
maintained in the posterior by default via the PMU, when the information provided is silent about correlations.

\begin{description}
\item[\textbf{DC2 Implementation:}]
\end{description}

Consider a composite system living in the Hilbert space $\mathcal{H}=\mathcal{H}%
_{1}\otimes \mathcal{H}_{2}$. Assume that all prior evidence led us to
believe the systems were independent. This is reflected in the prior
density matrix: if the individual system priors are $\hat{\varphi}_{1}$ and $%
\hat{\varphi}_{2}$, then the joint prior for the whole system is $%
\hat{\varphi}_{1}\otimes\hat{\varphi}_{2}$. Further suppose that new information is acquired
such that $\hat{\varphi}_{1}$ would by itself be updated to $\hat{\rho}_1$ and
that $\hat{\varphi}_{2}$ would be itself be updated to $\hat{\rho}_{2}$. By design, the implementation of DC2
constrains the entropy functional such that in this case, the joint product prior density matrix $%
\hat{\varphi}_{1}\otimes\hat{\varphi}_{2} $ updates to the product posterior $\hat{\rho}_{1}\otimes\hat{\rho}_{2} $
so that inferences about one do not affect inferences about the other.

The argument below is considerably simplified if we expand the space of
density matrices to include density matrices that are not necessarily normalized. This does
not represent any limitation because normalization can always be easily
achieved as one additional constraint. We consider a few special cases below: \\

\noindent \textbf{Case 1:} We receive the extremely constraining information
that the posterior distribution for system $1$ is completely specified to be 
$\hat{\rho}_1$ while we receive no information about system $2$ at all. We
treat the two systems jointly. Maximize the joint entropy $S[\hat{\rho}_{12},\hat{\varphi}_{1}\otimes\hat{\varphi}_{2} ]$, subject to the following constraints on the $\hat{\rho}_{12}\,$, 
\begin{equation}
\mbox{Tr}_2(\hat{\rho}_{12})=\hat{\rho}_{1}.
\end{equation}%
Notice all of the $N^2$ elements in $\mathcal{H}_1$ of $\hat{\rho}_{12}$ are being constrained. We therefore need a Lagrange multiplier which spans $\mathcal{H}_1$ and therefore it is a square matrix $\hat{\lambda}_1$. This is readily seen by observing the component form expressions of the Lagrange multipliers $(\hat{\lambda}_1)_{ij}=\lambda_{ij}$. Maximizing the entropy with respect to this $\mathcal{H}_2$ independent constraint is,
\begin{eqnarray}
0=\delta\Big(S-\sum_{ij}\lambda_{ij}\Big(\mbox{Tr}_2(\hat{\rho}_{1,2}) -\hat{\rho}_{1}\Big)_{ij}\Big),
\end{eqnarray}
but reexpressing this with its transpose $(\hat{\lambda}_1)_{ij}=(\hat{\lambda}_1^T)_{ji}$, gives
\begin{eqnarray}
0=\delta\Big(S-\mbox{Tr}_1(\hat{\lambda}_1[\mbox{Tr}_2(\hat{\rho}_{1,2}) -\hat{\rho}_{1}])\Big),
\end{eqnarray}
where we have relabeled $\hat{\lambda}_1^T\rightarrow \hat{\lambda}_1$, for convenience, as the name of the Lagrange multipliers are arbitrary. For arbitrary variations of $\hat{\rho}_{12}$, we therefore have,
\begin{equation}
\hat{\lambda}_1\otimes \hat{1}_2=\phi \left( \hat{\rho}_{12},\hat{\varphi}_{1}\otimes\hat{\varphi}_{2}\right)\,.
\end{equation}%
DC2 is implemented by requiring $\hat{\varphi}_{1}\otimes\hat{\varphi}_{2}\rightarrow \hat{\rho}_{1}\otimes\hat{\varphi}_{2}$, such that the function $\phi$ is designed to reflect subsystem independence in this case; therefore, we have
\begin{equation}
\hat{\lambda}_1\otimes \hat{1}_2=\phi \left( \hat{\rho}_{1}\otimes\hat{\varphi}_{2},\hat{\varphi}_{1}\otimes\hat{\varphi}_{2}\right).\label{qlambda1 a}
\end{equation}
This equation must hold for all choices of the independent prior $\hat{\varphi}_{2}$ in $\mathcal{H}_2$. Suppose we had chosen a different  prior $\hat{\varphi}_{2}^{\prime }=\hat{\varphi}_{2}+\delta \hat{\varphi}_{2}$. For all $\delta \hat{\varphi}_{2}$ the LHS $\hat{\lambda}_1\otimes \hat{1}_2$ remains unchanged. This means that any dependence that the right hand side might potentially have had on $\hat{\varphi}_{2}$ \emph{must cancel out}, meaning,
\begin{equation}
\phi \left( \hat{\rho}_{1}\otimes\hat{\varphi}_{2},\hat{\varphi}_{1}\otimes\hat{\varphi}_{2}\right)=f( \hat{\rho}_{1}, \hat{\varphi}_{1})\otimes\hat{1}_2.
\end{equation}
Since $\hat{\varphi}_{2}$ is arbitrary, suppose further that we choose a unit prior, $%
\hat{\varphi}_{2}=\hat{1}_2\,$, and note that $\hat{\rho}_{1}\otimes\hat{1}_2$ and $\hat{\varphi}_{1}\otimes\hat{1}_2$ are block diagonal in $\mathcal{H}_2$. Because the LHS is block diagonal in $\mathcal{H}_2$,
\begin{eqnarray}
f( \hat{\rho}_{1}, \hat{\varphi}_{1})\otimes\hat{1}_2=\phi \left( \hat{\rho}_{1}\otimes\hat{1}_{2},\hat{\varphi}_{1}\otimes\hat{1}_{2}\right)
\end{eqnarray}
the RHS is block diagonal in $\mathcal{H}_2$, and because the function $\phi$ is understood to be a power series expansion in its arguments, 
\begin{eqnarray}
f( \hat{\rho}_{1}, \hat{\varphi}_{1})\otimes\hat{1}_2=\phi \left( \hat{\rho}_{1}\otimes\hat{1}_{2},\hat{\varphi}_{1}\otimes\hat{1}_{2}\right)=\phi \left( \hat{\rho}_{1},\hat{\varphi}_{1}\right)\otimes\hat{1}_{2}.
\end{eqnarray}
This gives, 
\begin{equation}
\hat{\lambda}_1\otimes \hat{1}_2=\phi \left( \hat{\rho}_{1},\hat{\varphi}_{1}\right)\otimes\hat{1}_{2},  \label{qlambda1 b}
\end{equation}
and therefore the $\hat{1}_2$ factors out and $\hat{\lambda}_1=\phi \left( \hat{\rho}_{1},\hat{\varphi}_{1}\right)$. %
A similar argument exchanging systems $1$ and $2$ shows  $\hat{\lambda}_2=\phi \left( \hat{\rho}_{2},\hat{\varphi}_{2}\right)$ in this case. 

\noindent \textbf{Case 1 - Conclusion:} The analysis leads us to
conclude that when the system 2 is not updated the dependence on $\hat{\varphi}_{2}$ also
drops out,    
\begin{equation}
\phi \left( \hat{\rho}_{1}\otimes\hat{\varphi}_{2},\hat{\varphi}_{1}\otimes\hat{\varphi}_{2}\right)=\phi \left( \hat{\rho}_{1},\hat{\varphi}_{1}\right)\otimes\hat{1}_2\label{qlambda1 c},
\end{equation} 
and similarly,
\begin{equation}
\phi \left( \hat{\varphi}_{1}\otimes\hat{\rho}_{2},\hat{\varphi}_{1}\otimes\hat{\varphi}_{2}\right)=\hat{1}_1\otimes\phi \left( \hat{\rho}_{2},\hat{\varphi}_{2}\right)\label{qlambda1 d}.
\end{equation} 

\noindent \textbf{Case 2:} Now consider a different special case in which
the marginal posterior distributions for systems $1$ and $2$ are both
completely specified to be $\hat{\rho}_1$ and $\hat{\rho}_2$ respectively.
Maximize the joint entropy, $S[\hat{\rho}_{12},\hat{\varphi}_{1}\otimes\hat{\varphi}_{2} ]$,
subject to the following constraints on the $\hat{\rho}_{12}\,$, 
\begin{equation}
\mbox{Tr}_2(\hat{\rho}_{12})=\hat{\rho}_{1}\quad \mbox{and}\quad \mbox{Tr}_1(\hat{\rho}_{12})=\hat{\rho}_{2}.
\end{equation}%
Here each expectation value constraints the entire space $\mathcal{H}_i$, where $\hat{\rho}_{i}$ lives. The Lagrange multipliers must span their respective spaces, so we implement the constraint with the Lagrange multiplier operator $\hat{\mu}_i$, then,
\begin{eqnarray}
0=\delta\Big(S-\mbox{Tr}_1(\hat{\mu}_1[\mbox{Tr}_2(\hat{\rho}_{12}) -\hat{\rho}_{1}])-\mbox{Tr}_2(\hat{\mu}_2[\mbox{Tr}_1(\hat{\rho}_{12}) -\hat{\rho}_{2}])\Big).
\end{eqnarray}
For arbitrary variations of $\hat{\rho}_{12}$, we have,
\begin{equation}
\hat{\mu}_1\otimes \hat{1}_2+\hat{1}_1\otimes \hat{\mu}_2=\phi \left( \hat{\rho}_{12},\hat{\varphi}_{1}\otimes\hat{\varphi}_{2}\right)\,.
\end{equation}%
By design, DC2 is implemented by requiring $\hat{\varphi}_{1}\otimes\hat{\varphi}_{2}\rightarrow \hat{\rho}_{1}\otimes\hat{\rho}_{2}$ in this case; therefore, we have
\begin{equation}
\hat{\mu}_1\otimes \hat{1}_2+\hat{1}_1\otimes \hat{\mu}_2=\phi \left( \hat{\rho}_{1}\otimes\hat{\rho}_{2},\hat{\varphi}_{1}\otimes\hat{\varphi}_{2}\right)\,\label{qmu12 a}.
\end{equation}
 Write (\ref{qmu12 a}) as,  
\begin{equation}
\hat{\mu}_1\otimes \hat{1}_2=\phi \left( \hat{\rho}_{1}\otimes\hat{\rho}_{2},\hat{\varphi}_{1}\otimes\hat{\varphi}_{2}\right)-\hat{1}_1\otimes \hat{\mu}_2 ~.  \label{1mu12 b}
\end{equation}%
The left hand side is independent of changes in of $\hat{\rho}_{2}$ and $\hat{\varphi}_{2}$ in $\mathcal{H}_2$ as $\hat{\mu}_2$ ``pushes out" this dependence from $\phi$.  Any dependence that the RHS might potentially have had on $\hat{\rho}_{2}$, $\hat{\varphi}_{2}$ must cancel out, leaving $\hat{\mu}_{1}$ unchanged. Consequently,
\begin{equation}
\phi \left(
\hat{\rho}_1\otimes \hat{\rho}_2,\hat{\varphi}_{1}\otimes \hat{\varphi}_{2}\right)-\hat{1}_1\otimes\hat{\mu}_{2}=g(\hat{\rho}_{1},\hat{\varphi}_{1})\otimes \hat{1}_2.
\end{equation}
Because $g(\hat{\rho}_{1},\hat{\varphi}_{1})$ is independent of arbitrary variations of $\hat{\rho}_{2}$ and $\hat{\varphi}_{2}$ on the LHS above -- it is satisfied equally well for all choices.  The form of $g(\hat{\rho}_{1},\hat{\varphi}_{1})$ reduces to the form of $f(\hat{\rho}_{1},\hat{\varphi}_{1})$ from Case 1 when $\hat{\rho}_{2}=\hat{\varphi}_{2}=\hat{1}_2$ and similarly DC1' gives $\hat{\mu}_2=0$. Therefore, the Lagrange multiplier is 
\begin{equation}
\hat{\mu}_1\otimes \hat{1}_2=\phi(\hat{\rho}_{1},\hat{\varphi}_{1})\otimes \hat{1}_2.
\end{equation}%
A similar analysis can be carried out for $\hat{\mu}_{2}$ leading to  
\begin{equation}
\hat{1}_1\otimes \hat{\mu}_2=\hat{1}_1\otimes \phi(\hat{\rho}_{2},\hat{\varphi}_{2}).
\end{equation}
\noindent \textbf{Case 2 - Conclusion:}
Substituting back into (\ref{qmu12 a}) gives us a functional equation for $%
\phi \,$,     
\begin{eqnarray}
\phi(\hat{\rho}_{1}\otimes\hat{\rho}_2,\hat{\varphi}_1\otimes\hat{\varphi}_2)=\phi(\hat{\rho}_{1},\hat{\varphi}_1)\otimes\hat{1}_2+\hat{1}_1\otimes\phi(\hat{\rho}_2,\hat{\varphi}_2),\label{77}
\end{eqnarray}
which is,
\begin{eqnarray}
\phi(\hat{\rho}_{1}\otimes\hat{\rho}_2,\hat{\varphi}_1\otimes\hat{\varphi}_2)=\phi(\hat{\rho}_{1}\otimes\hat{1}_2,\hat{\varphi}_1\otimes\hat{1}_2)+\phi(\hat{1}_1\otimes\hat{\rho}_2,\hat{1}_1\otimes\hat{\varphi}_2).\label{55}
\end{eqnarray}
The general solution to this matrix valued functional equation is derived in the Appendix \ref{appendix2}, and is,
\begin{eqnarray}
\phi(\hat{\rho},\hat{\varphi})=\stackrel{\,\,\sim}{A}\ln(\hat{\rho})+\stackrel{\,\,\sim}{B}\ln(\hat{\varphi}),
\end{eqnarray}
where tilde $\stackrel{\,\,\sim}{A}$ is a ``super-operator" having constant coefficients and twice the number of indicies as $\hat{\rho}$ and $\hat{\varphi}$ as discussed in the Appendix (i.e. $\left(\stackrel{\,\,\sim}{A}\ln(\hat{\rho})\right)_{ij}=\sum_{k\ell}A_{ijk\ell}(\log(\hat{\rho}))_{k\ell}$ and similarly for $\stackrel{\,\,\sim}{B}\ln(\hat{\varphi})$). DC1' imposes,
\begin{eqnarray}
\phi(\hat{\varphi},\hat{\varphi})=\stackrel{\,\,\sim}{A}\ln(\hat{\varphi})+\stackrel{\,\,\sim}{B}\ln(\hat{\varphi})=\hat{0},
\end{eqnarray}
which is satisfied in general when $\stackrel{\,\,\sim}{A}=-\stackrel{\,\,\sim}{B}$, and now,
\begin{eqnarray}
\phi(\hat{\rho},\hat{\varphi})=\stackrel{\,\,\sim}{A}\Big(\ln(\hat{\rho})-\ln(\hat{\varphi})\Big).
\end{eqnarray}
We may fix the constant $\stackrel{\,\,\sim}{A}$ by substituting our solution into the RHS of equation (\ref{77}) which is equal to the RHS of equation (\ref{55}),
\begin{eqnarray}
\Big(\stackrel{\,\,\sim}{A}_{1}\Big(\ln(\hat{\rho}_{1})-\ln(\hat{\varphi}_1)\Big)\Big)\otimes\hat{1}_2+\hat{1}_1\otimes\Big(\stackrel{\,\,\sim}{A}_{2}\Big(\ln(\hat{\rho}_{2})-\ln(\hat{\varphi}_2)\Big)\Big)\nonumber
\end{eqnarray}
\begin{eqnarray}
=\stackrel{\,\,\sim}{A}_{12}\Big(\ln(\hat{\rho}_{1}\otimes\hat{1}_2)-\ln(\hat{\varphi}_1\otimes\hat{1}_2)\Big)+\stackrel{\,\,\sim}{A}_{12}\Big(\ln(\hat{1}_{1}\otimes\hat{\rho}_2)-\ln(\hat{1}_1\otimes\hat{\varphi}_2)\Big),\label{59}
\end{eqnarray}
where $\stackrel{\,\,\sim}{A}_{12}$ acts on the joint space of $1$ and $2$ and $\stackrel{\,\,\sim}{A}_{1}$, $\stackrel{\,\,\sim}{A}_{2}$ acts on single subspaces $1$ or $2$ respectively. Using the log tensor product identity, $\ln(\hat{\rho}_{1}\otimes\hat{1}_2)=\ln(\hat{\rho}_{1})\otimes\hat{1}_2$, in the RHS of equation (\ref{59}) gives,
\begin{eqnarray}
=\stackrel{\,\,\sim}{A}_{12}\Big(\ln(\hat{\rho}_{1})\otimes\hat{1}_2-\ln(\hat{\varphi}_1)\otimes\hat{1}_2\Big)+\stackrel{\,\,\sim}{A}_{12}\Big(\hat{1}_{1}\otimes\ln(\hat{\rho}_2)-\hat{1}_1\otimes\ln(\hat{\varphi}_2)\Big).\label{60}
\end{eqnarray}
Note that arbitrarily letting $\hat{\rho}_2=\hat{\varphi}_2$ gives,
\begin{eqnarray}
\Big(\stackrel{\,\,\sim}{A}_{1}\Big(\ln(\hat{\rho}_{1})-\ln(\hat{\varphi}_1)\Big)\Big)\otimes\hat{1}_2
=\stackrel{\,\,\sim}{A}_{12}\Big(\ln(\hat{\rho}_{1})\otimes\hat{1}_2-\ln(\hat{\varphi}_1)\otimes\hat{1}_2\Big).
\end{eqnarray}
or arbitrarily letting $\hat{\rho}_1=\hat{\varphi}_1$ gives,
\begin{eqnarray}
\hat{1}_1\otimes\Big(\stackrel{\,\,\sim}{A}_{2}\Big(\ln(\hat{\rho}_{2})-\ln(\hat{\varphi}_2)\Big)\Big)=\stackrel{\,\,\sim}{A}_{12}\Big(\hat{1}_{1}\otimes\ln(\hat{\rho}_2)-\hat{1}_1\otimes\ln(\hat{\varphi}_2)\Big).
\end{eqnarray}
As $\stackrel{\,\,\sim}{A}_{12}$, $\stackrel{\,\,\sim}{A}_{1}$, and $\stackrel{\,\,\sim}{A}_{2}$ are constant tensors, inspecting the above equalities determines the form of the tensor to be $\stackrel{\,\,\sim}{A}\,=\stackrel{\,\,}{A}\stackrel{\,\,\sim}{1}$ where $A$ is a scalar constant and $\stackrel{\,\,\sim}{1}$ is the super-operator identity over the appropriate (joint) Hilbert space.

Because our goal is to maximize the entropy function, we let the arbitrary constant $A=-|A|$ and distribute $\stackrel{\,\,\sim}{1}$ identically, which gives the final functional form,
\begin{eqnarray}
\phi(\hat{\rho},\hat{\varphi})=-|A|\Big(\ln(\hat{\rho})-\ln(\hat{\varphi})\Big).
\end{eqnarray}
 ``Integrating" $\phi$, gives a general form for the quantum relative entropy,
\begin{eqnarray}
S(\hat{\rho},\hat{\varphi} )=-|A|\mbox{Tr}(\hat{\rho} \log \hat{\rho} -\hat{\rho}\log \hat{\varphi} -\hat{\rho})+C[\hat{\varphi}]=-|A|S_U(\hat{\rho},\hat{\varphi})+|A|\mbox{Tr}(\hat{\rho})+C[\hat{\varphi}], 
\end{eqnarray}
where $S_U(\hat{\rho},\hat{\varphi})$ is Umegaki's form of the relative entropy, the extra $|A|\mbox{Tr}(\hat{\rho})$ from integration is an artifact present for the preservation of DC1', and $C[\hat{\varphi}]$ is a constant in the sense that it drops out under arbitrary variations of $\hat{\rho}$. This entropy leads to the same inferences as Umegaki's form of the entropy with added bonus that $\hat{\rho}=\hat{\varphi}$ in the absence of constraints or changes in information -- rather than $\hat{\rho}=e^{-1}\hat{\varphi}$ which would be given by maximizing Umegaki's form of the entropy. In this sense the extra $|A|\mbox{Tr}(\hat{\rho})$ only improves the inference process as it more readily adheres to the PMU though DC1'; however now because $S_{U}\geq 0$, we have $S(\hat{\rho},\hat{\varphi})\leq \mbox{Tr}(\hat{\rho})+C[\hat{\varphi}]$, which provides little nuisance. In the spirit of this derivation we will keep the $\mbox{Tr}(\hat{\rho})$ term there, but for all practical purposes of inference, as long as there is a normalization constraint, it plays no role, and we find (letting $|A|=1$ and $C[\hat{\varphi}]=0$), 
\begin{eqnarray}
S(\hat{\rho},\hat{\varphi} )\rightarrow S^*(\hat{\rho},\hat{\varphi} )=-S_U(\hat{\rho},\hat{\varphi})=-\mbox{Tr}(\hat{\rho} \log \hat{\rho} -\hat{\rho}\log \hat{\varphi}),
\end{eqnarray}
Umegaki's form of the relative entropy. $S^*(\hat{\rho},\hat{\varphi})$ is an equally valid entropy because, given normalization is applied, the same selected posterior $\hat{\rho}$ maximizes both $S(\hat{\rho},\hat{\varphi})$ and $S^*(\hat{\rho},\hat{\varphi})$.

\subsection{Remarks}
 Due to the universality and the equal application of the PMU by using the same design criteria for both the standard and quantum case, the quantum relative entropy reduces to the standard relative entropy when $[\hat{\rho},\hat{\varphi}]=0$ or when the experiment being preformed $\hat{\rho}\rightarrow \rho(a)=\mbox{Tr}(\hat{\rho}\ket{a}\bra{a})$ is known. The quantum relative entropy we derive has the correct asymptotic form of the standard relative entropy in the sense of \cite{Petz1,Petz2,Petz3}. Further connections will be illustrated in a follow up article that is concerned with direct applications of the quantum relative entropy. Because two entropies are derived in parallel, we expect the well known inferential results and consequences of the relative entropy to have a quantum relative entropy representation. 

Maximizing the quantum relative entropy with respect to some constraints $\expt{\hat{A}_i}$, where $\{\hat{A}_i\}$ are a set of  arbitrary Hermitian operators, and normalization $\expt{\hat{1}}=1$, gives the following general solution for the posterior density matrix:
\begin{eqnarray}
\hat{\rho}=\exp\Big(\alpha_0\hat{1}+\sum_i\alpha_i\hat{A}_i+\ln(\hat{\varphi})\Big)=\frac{1}{Z}\exp\Big(\sum_i\alpha_i\hat{A}_i+\ln(\hat{\varphi})\Big)\equiv \frac{1}{Z}\exp\Big(\hat{C}\Big),\label{result}
\end{eqnarray}
where $\alpha_i$ are the Lagrange multipliers of the respective constraints and normalization may be factored out of the exponential in general because the identity commutes universally. If $\hat{\varphi}\propto \hat{1}$, it is well known the analysis arrives at the same expression for $\hat{\rho}$ after normalization as it would if the von Neumann entropy were used, and thus one can find expressions for thermalized quantum states $\hat{\rho}=\frac{1}{Z}e^{-\beta \hat{H}}$.   The remaining problem is to solve for the $N$ Lagrange multipliers using their $N$ associated expectation value constraints. In principle their solution is found by computing $Z$ and using standard methods from Statistical Mechanics,
\begin{eqnarray}
\expt{\hat{A}_i}=-\frac{\d}{\d \alpha_i}\ln(Z)\label{gen},
\end{eqnarray}
and inverting to find $\alpha_i=\alpha_i(\expt{\hat{A}_i})$, which has a unique solution due to the joint concavity (convexity depending on the sign convention) of the quantum relative entropy \cite{Petz1, Petz2} when the constraints are linear in $\hat{\rho}$. Between the Zassenhaus formula
\begin{eqnarray}
e^{t(\hat{A}+\hat{B})}=e^{t\hat{A}}e^{t\hat{B}}e^{-\frac{t^2}{2}[\hat{A},\hat{B}]}e^{\frac{t^3}{6}(2[\hat{B},[\hat{A},\hat{B}]]+[\hat{A},[\hat{A},\hat{B}]])}...,
\end{eqnarray}
and Horn's inequality, the solutions to (\ref{gen}) lack a certain calculational elegance because it is difficult to express the eigenvalues of $\hat{C}=\log(\hat{\varphi})+\sum\alpha_i\hat{A}_i$ (in the exponential) in simple terms of the eigenvalues of the $\hat{A}_i$'s and $\hat{\varphi}$, in general, when the matrices do not commute. The solution requires solving the eigenvalue problem for $\hat{C}$, such the the exponential of $\hat{C}$ may be taken and evaluated in terms of the eigenvalues of the $\alpha_i\hat{A}_i$'s and the prior density matrix $\hat{\varphi}$.  A pedagogical exercise is, starting with a prior which is a mixture of spin-z up and down $\hat{\varphi}=a\ket{+}\bra{+}+b\ket{-}\bra{-}$ ($a,b\neq0$) and maximize the quantum relative entropy with respect to the expectation of a general Hermitian operator. This example is given in the Appendix \ref{spinexample}.

\section{Conclusions:}

This approach emphasizes the notion that entropy is a tool for performing 
inference and downplays counter-notional issues which arise if one interprets entropy as a measure
of disorder, a measure of distinguishability, or an amount of missing information \cite{book}. Because the same design criteria, guided by the PMU, are applied equally well to the design of a relative and quantum relative entropy, we find that both the relative and quantum relative entropy are designed for the purpose of inference. Because the quantum relative entropy is the function which fits the requirements of a tool designed for inference, we now know what it is and how to use it -- formulating an inferential quantum maximum entropy method. A follow up article is concerned with a few interesting applications of the quantum maximum entropy method, and in particular it derives the Quantum Bayes Rule. \newpage
\section{Acknowledgments}
 I must give ample acknowledgment to Ariel Caticha who suggested the problem of justifying the form of the quantum relative entropy as a criterion for ranking of density matrices. He cleared up several difficulties by suggesting that design constraints be applied to the variational derivative of the entropy rather than the entropy itself. As well, he provided substantial improvements to the method for imposing DC2 that lead to the functional equations for the variational derivatives ($\phi_{12}=\phi_{1}+\phi_{2}$) -- with more rigor than in earlier versions of this article. His time and guidance are all greatly appreciated -- Thanks Ariel.

\section{Appendix:}
The Appendix loosely follows the relevant sections in \cite{Aczel}, and then uses the methods reviewed to solve the relevant functional equations for $\phi$. The last section is an example of the quantum maximum entropy method for spin.
\subsection{Simple functional equations}
 From \cite{Aczel} pages 31-44.

\paragraph{Thm 1:} \emph{If Cauchy's functional equation
\begin{eqnarray}
f(x+y)=f(x)+f(y),\label{Cauchy}
\end{eqnarray}
is satisfied for all real $x$, $y$, and if the function $f(x)$ is (a) continuous at a point, (b) nonegative for small positive $x$'s, or (c) bounded in an interval, then,
\begin{eqnarray}
f(x)=cx\label{linear}
\end{eqnarray}
is the solution to (\ref{Cauchy}) for all real $x$. If (\ref{Cauchy}) is assumed only over all positive $x$, $y$, then under the same conditions (\ref{linear}) holds for all positive $x$.} 
\paragraph{Proof 1:}
The most natural assumption for our purposes is that $f(x)$ is continuous at a point (which later extends to continuity all points as given by Darboux). Cauchy solved the functional equation by induction. In particular equation (\ref{Cauchy}) implies,
\begin{eqnarray}
f(\sum_{i}x_i)=\sum_if(x_i),
\end{eqnarray}
and if we let each $x_i=x$ as a special case to determine $f$, we find 
\begin{eqnarray}
f(nx)=nf(x).
\end{eqnarray}
We may let $nx=mt$ such that
\begin{eqnarray}
f(x)=f(\frac{m}{n}t)=\frac{m}{n}f(t).
\end{eqnarray}
Letting $\lim_{t\rightarrow 1}f(t)=f(1)=c$, gives
\begin{eqnarray}
f(\frac{m}{n})=\frac{m}{n}f(1)=\frac{m}{n}c,
\end{eqnarray}
and because for $t=1$,  $x=\frac{m}{n}$ above, we have
\begin{eqnarray}
f(x)=cx,
\end{eqnarray}
which is the general solution of the linear functional equation. In principle $c$ can be complex. The importance of Cauchy's solution is that can be used to give general solutions to the following Cauchy equations:
\begin{eqnarray}
f(x+y)&=&f(x)f(y),\\
f(xy)&=&f(x)+f(y),\\
f(xy)&=&f(x)f(y),
\end{eqnarray}
by preforming consistent substitution until they are the same form as (\ref{Cauchy}) as given by Cauchy. We will briefly discuss the first two.
\paragraph{Thm 2:}\emph{ The general solution of $f(x+y)=f(x)f(y)$ is $f(x)=e^{cx}$ for all real or for all positive $x,y$ that are continuous at one point and, in addition to the exponential solution, the solution $f(0)=1$ and $f(x)=0$ for ($x>0$) are in these classes of functions.}

The first functional $f(x+y)=f(x)f(y)$ is solved by first noting that it is strictly positive for real $x$, $y$, $f(x)$, which can be shown by considering $x=y$,
\begin{eqnarray}
f(2x)=f(x)^2> 0.
\end{eqnarray}
If there exists $f(x_0)=0$, then it follows that $f(x)=f((x-x_0)+x_0)=0$, a trivial solution, hence why the possibility of being equal to zero is excluded above. Given $f(x)$ is nowhere zero, we are justified in taking the natural logarithm $\ln(x)$, due to its positivity $f(x)>0$. This gives,
\begin{eqnarray}
\ln(f(x+y))=\ln(f(x))+\ln(f(y)),
\end{eqnarray}
and letting $g(x)=\ln(f(x))$ gives,
\begin{eqnarray}
g(x+y)=g(x)+g(y),
\end{eqnarray}
which is Cauchy's linear equation, and thus has the solution $g(x)=cx$. Because  $g(x)=\ln(f(x))$, one finds in general that $f(x)=e^{cx}$.
\paragraph{Thm 3:}\emph{ If the functional equation $f(xy)=f(x)+f(y)$ is valid for all positive $x,y$ then its general solution is $f(x)=c\ln(x)$ given it is continuous at a point. If $x=0$ (or $y=0$) are valid then the general solution is $f(x)=0$. If all real $x,y$ are valid except $0$ then the general solution is $f(x)=c\ln(|x|)$. }

 In particular we are interested in the functional equation $f(xy)=f(x)+f(y)$ when $x,y$ are positive. In this case we can again follow Cauchy and substitute $x=e^u$ and $y=e^v$ to get,
\begin{eqnarray}
f(e^ue^v)=f(e^u)+f(e^v),\label{87}
\end{eqnarray}
and letting $g(u)=f(e^u)$ gives $g(u+v)=g(u)+g(v)$. Again, the solution is $g(u)=cu$ and therefore the general solution is $f(x)=c\ln(x)$ when we substitute for $u$. If $x$ could equal $0$ then $f(0)=f(x)+f(0)$, which has the trivial solution $f(x)=0$. The general solution for $x\neq0$, $y\neq0$ and $x,y$ positive is therefore $f(x)=c\ln(x)$. 

\subsection{Functional equations with multiple arguments}
From \cite{Aczel} pages 213-217. Consider the functional equation,
\begin{eqnarray}
F(x_1+y_1,x_2+y_2,...,x_n+y_n)=F(x_1,x_2,...,x_n)+F(y_1,y_2,...,y_n),\label{GenCauchy}
\end{eqnarray}
which is a generalization of Cauchy's linear functional equation (\ref{Cauchy}) to several arguments. Letting $x_2=x_3=...=x_n=y_2=y_3=...=y_n=0$ gives
\begin{eqnarray}
F(x_1+y_1,0,...,0)=F(x_1,0,...,0)+F(y_1,0,...,0),
\end{eqnarray}
which is the Cauchy linear functional equation having solution $F(x_1,0,...,0)=c_1x_1$ where $F(x_1,0,...,0)$ is assumed to be continuous or at least measurable majorant. Similarly,
\begin{eqnarray}
F(0,...,0,x_k,0,...,0)=c_kx_k,
\end{eqnarray}
and if you consider
\begin{eqnarray}
F(x_1+0,0+y_2,0,...,0)=F(x_1,0,...,0)+F(0,y_2,0,...,0)=c_1x_1+c_2y_2,
\end{eqnarray}
and as $y_2$ is arbitrary we could have let $y_2=x_2$ such that in general
\begin{eqnarray}
F(x_1,x_2,...,x_n)=\sum c_ix_i,\label{genlinearsol}
\end{eqnarray}
as a general solution.
\subsection{Relative entropy\label{appendix1}:}
We are interested in the following functional equation,
\begin{eqnarray}
\phi(\rho_1\rho_2,\varphi_1\varphi_2)=\phi(\rho_1,\varphi_1)+\phi(\rho_2,\varphi_2).
\end{eqnarray}
This is an equation of the form,
\begin{eqnarray}
F(x_1y_1,x_2y_2)=F(x_1,x_2)+F(y_1,y_2),
\end{eqnarray}
where $x_1=\rho(x_1)$, $y_1=\rho(x_2)$, $x_2=\varphi(x_1)$, and $y_2=\varphi(x_2)$. First assume all $q$ and $p$ are greater than zero. Then, substitute: $x_i=e^{x_i'}$ and $y_i=e^{y_i'}$ and let $F'(x_1',x_2')=F(e^{x_1'},e^{x_2'})$ and so on such that
\begin{eqnarray}
F'(x_1'+y_1',x_2'+y_2')=F'(x_1',x_2')+F'(y_1',y_2'),
\end{eqnarray}
which is of the form of (\ref{GenCauchy}). The general solution for $F$ is therefore
\begin{eqnarray}
F'(x_1'+y_1',x_2'+y_2')=a_1(x_1'+y_1')+a_2(x_2'+y_2')=a_1\ln(x_1y_1)+a_2\ln(x_2y_2)=F(x_1y_1,x_2y_2)
\end{eqnarray}
which means the general solution for $\phi$ is,
\begin{eqnarray}
\phi(\rho_1,\varphi_1)&=&a_1\ln(\rho(x_1))+a_2\ln(\varphi(x_1))
\end{eqnarray}
In such a case when $\varphi(x_0)=0$ for some value $x_0\in \mathcal{X}$ we may let $\varphi(x_0)=\epsilon$ where $\epsilon$ is as close to zero as we could possibly want -- the trivial general solution $\phi=0$ is saturated by the special case when $\rho=\varphi$ from DC1'. Here we return to the text.

\subsection{Matrix functional equations}
(This derivation is implied in \cite{Aczel} pages 347-349). First consider a Cauchy matrix functional equation,
\begin{eqnarray}
f(\hat{X}+\hat{Y})=f(\hat{X})+f(\hat{Y})\label{102}
\end{eqnarray}
where $\hat{X}$ and $\hat{Y}$ are $n\times n$ square matrices. Rewriting the matrix functional equation in terms of its components gives,
\begin{eqnarray}
f_{ij}(x_{11}+y_{11},x_{12}+y_{12},...,x_{nn}+y_{nn})=f_{ij}(x_{11},x_{12},...,x_{nn})+f_{ij}(y_{11},y_{12},...,y_{nn})
\end{eqnarray}
is now in the form of (\ref{GenCauchy}) and therefore the solution is,
\begin{eqnarray}
f_{ij}(x_{11},x_{12},...,x_{nn})=\sum_{\ell,k=0}^n c_{ij\ell k}x_{\ell k}
\end{eqnarray}
for $i,j=1,...,n$. We find it convenient to introduce super indices, $A=(i,j)$ and $B=(\ell, k)$ such that the component equation becomes,
\begin{eqnarray}
f_{A}=\sum_{B} c_{AB}x_{B}.\label{cab}
\end{eqnarray}
resembles the solution for a linear transformation of a vector from \cite{Aczel}. In general we will be discussing matrices $\hat{X}=\hat{X}_1\otimes\hat{X}_2\otimes...\otimes\hat{X}_N$ which stem out of the tensor products of density matrices. In this situation $\hat{X}$ can be thought of as $2N$ index tensor or a $z\times z$ matrix where $z=\prod_{i}^N n_i$ is the product of the ranks of the matrices in the tensor product or even $\hat{X}$ is a vector of length $z^2$. In such a case we may abuse the super index notation where $A$ and $B$ lump together the appropriate number of indices such that (\ref{cab}) is the form of the solution for the components in general. The matrix form of the general solution is,
\begin{eqnarray}
f(\hat{X})=\widetilde{C}\hat{X},
\end{eqnarray}
where $\widetilde{C}$ is a constant super-operator having components $c_{AB}$.
\subsection{Quantum Relative entropy:\label{appendix2}} The functional equation is,
\begin{eqnarray}
\phi\Big(\hat{\rho}_{1}\otimes\hat{\rho}_2,\hat{\varphi}_1\otimes\hat{\varphi}_2\Big)=\phi\Big(\hat{\rho}_{1}\otimes\hat{1}_2,\hat{\varphi}_1\otimes\hat{1}_2\Big)+\phi\Big(\hat{1}_1\otimes\hat{\rho}_2,\hat{1}_1\otimes\hat{\varphi}_2\Big).
\end{eqnarray}
These density matrices are Hermitian, positive semi-definite, have positive eigenvalues, and are not equal to $\hat{0}$. Because every invertible matrix can be expressed as the exponential of some other matrix, we can substitute $\hat{\rho}_{1}=e^{\hat{\rho}_{1}'}$, and so on for all four density matrices which gives,
\begin{eqnarray}
\phi\Big( e^{\hat{\rho}_{1}'}\otimes e^{\hat{\rho}_2'},e^{\hat{\varphi}_1'}\otimes e^{\hat{\varphi}_2'}\Big)=\phi\Big(e^{\hat{\rho}_{1}'}\otimes\hat{1}_2,e^{\hat{\varphi}_1'}\otimes\hat{1}_2\Big)+\phi\Big(\hat{1}_1\otimes e^{\hat{\rho}_2'},\hat{1}_1\otimes e^{\hat{\varphi}_2'}\Big).
\end{eqnarray}
Now we use the following identities for Hermitian matrices, 
\begin{eqnarray}
 e^{\hat{\rho}_{1}'}\otimes e^{\hat{\rho}_2'}=e^{\hat{\rho}_{1}'\otimes\hat{1}_2+ \hat{1}_1\otimes\hat{\rho}_2'}
\end{eqnarray}
and 
\begin{eqnarray}
 e^{\hat{\rho}_{1}'}\otimes \hat{1_2}=e^{\hat{\rho}_{1}'\otimes\hat{1}_2},
\end{eqnarray}
to recast the functional equation as,
\begin{eqnarray}
\phi\Big(e^{\hat{\rho}_{1}'\otimes\hat{1}_2+ \hat{1}_1\otimes\hat{\rho}_2'},e^{\hat{\varphi}_{1}'\otimes\hat{1}_2+ \hat{1}_1\otimes\hat{\varphi}_2'}\Big)=\phi\Big(e^{\hat{\rho}_{1}'\otimes\hat{1}_2},e^{\hat{\varphi}_{1}'\otimes\hat{1}_2}\Big)+\phi\Big(e^{\hat{1}_1\otimes \hat{\rho}_2'},e^{\hat{1}_1\otimes \hat{\varphi}_2'}\Big).
\end{eqnarray}
Letting $G(\hat{\rho}_{1}'\otimes\hat{1}_2,\hat{\varphi}_{1}'\otimes\hat{1}_2)=\phi\Big(e^{\hat{\rho}_{1}'\otimes\hat{1}_2},e^{\hat{\varphi}_{1}'\otimes\hat{1}_2}\Big)$ gives,
\begin{eqnarray}
G(\hat{\rho}_{1}'\otimes\hat{1}_2+ \hat{1}_1\otimes\hat{\rho}_2',\hat{\varphi}_{1}'\otimes\hat{1}_2+ \hat{1}_1\otimes\hat{\varphi}_2')
=G(\hat{\rho}_{1}'\otimes\hat{1}_2,\hat{\varphi}_{1}'\otimes\hat{1}_2)+G(\hat{1}_1\otimes\hat{\rho}_2',\hat{1}_1\otimes\hat{\varphi}_2').
\end{eqnarray}
This functional equation is of the form
\begin{eqnarray}
G(\hat{X}_1'+\hat{Y}_1',\hat{X}_2'+\hat{Y}_2')=G(\hat{X}_1',\hat{X}_2')+G(\hat{Y}_1',\hat{Y}_2'),
\end{eqnarray}
which has the general solution
\begin{eqnarray}
G(\hat{X}',\hat{Y}')=\stackrel{\,\,\sim}{A}\hat{X}'+\widetilde{B}\hat{Y}',
\end{eqnarray}
synonymous to (\ref{genlinearsol}), and finally in general,
\begin{eqnarray}
\phi(\hat{\rho},\hat{\varphi})=\stackrel{\,\,\sim}{A}\ln(\hat{\rho})+\widetilde{B}\ln(\hat{\varphi}).
\end{eqnarray}
where $\stackrel{\,\,\sim}{A},\stackrel{\,\,\sim}{B}$ are super-operators having constant coefficients. 

\subsection{Spin Example\label{spinexample}}
Consider an arbitrarily mixed prior is (in the spin-$z$ basis for convenience) with $a,b\neq0$,
\begin{eqnarray}
\hat{\varphi}=a\ket{+}\bra{+}+b\ket{-}\bra{-}
\end{eqnarray}
and a general Hermitian matrix in the spin-$1/2$ Hilbert space, 
\begin{eqnarray}
c_{\mu}\hat{\sigma}^{\mu}=c_1\hat{1}+c_x\hat{\sigma}_x+c_y\hat{\sigma}_x+c_z\hat{\sigma}_z
\end{eqnarray}
\begin{eqnarray}
=(c_1+c_z)\ket{+}\bra{+}+(c_x-ic_y)\ket{+}\bra{-}+(c_x+ic_y)\ket{-}\bra{+}+(c_1-c_z)\ket{-}\bra{-},
\end{eqnarray}
having a known expectation value,
\begin{eqnarray}
\mbox{Tr}(\hat{\rho}c_{\mu}\hat{\sigma}^{\mu})=c.
\end{eqnarray}
Maximizing the entropy with respect to this general expectation value and normalization is:
\begin{eqnarray}
0=\Big(\delta S-\lambda[\mbox{Tr}(\hat{\rho})-1]-\alpha(\mbox{Tr}(\hat{\rho}c_{\mu}\hat{\sigma}^{\mu})-c)\Big),
\end{eqnarray}
which after varying gives,
\begin{eqnarray}
\hat{\rho}=\frac{1}{Z}\exp(\alpha c_{\mu}\hat{\sigma}^{\mu}+\log(\hat{\varphi}))\label{spinrhoexample}.
\end{eqnarray}
Letting
\begin{eqnarray}
\hat{C}=\alpha c_{\mu}\hat{\sigma}^{\mu}+\log(\hat{\varphi})
\end{eqnarray}
gives
\begin{eqnarray}
\hat{\rho}=\frac{1}{Z}e^{\hat{C}}=Ue^{U^{-1}\hat{C}U}U^{-1}=\frac{1}{Z}Ue^{\hat{\lambda}}U^{-1}\nonumber\\
=\frac{e^{\lambda_+}}{Z}U\ket{\lambda_+}\bra{\lambda_+}U^{-1}+\frac{e^{\lambda_-}}{Z}U\ket{\lambda_-}\bra{\lambda_-}U^{-1},
\end{eqnarray}
where $\hat{\lambda}$ is the diagonalized matrix of $\hat{C}$ having the real eigenvalues. They are,
\begin{eqnarray}
\lambda_{\pm}=\lambda\pm\delta\lambda,
\end{eqnarray}
due to the quadratic formula, explicitly:
\begin{eqnarray}
\lambda=\alpha c_1+\frac{1}{2}\log(ab),
\end{eqnarray}
and
\begin{eqnarray}
\delta\lambda=\frac{1}{2}\sqrt{\Big(2\alpha c_z+\log(\frac{a}{b})\Big)^2+4\alpha^2(c_x^2+c_y^2)}.
\end{eqnarray}
Because $\lambda_{\pm}$ and $a,b,c_1,c_x,c_y,c_z$ are real, $\delta\lambda\geq 0$.
The normalization constraint specifies the Lagrange multiplier $Z$,
\begin{eqnarray}
1=\mbox{Tr}(\hat{\rho})=\frac{e^{\lambda_+}+e^{\lambda_-}}{Z},
\end{eqnarray}
so $Z=e^{\lambda_+}+e^{\lambda_-}=2e^{\lambda}\cosh(\delta\lambda)$. The expectation value constraint specifies the Lagrange multiplier $\alpha$,
\begin{eqnarray}
c=\mbox{Tr}(\hat{\rho}c_{\mu}\sigma^{\mu})=\frac{\d}{\d\alpha}\log(Z)=c_1+\tanh(\delta\lambda)\frac{\d }{\d\alpha}\delta\lambda,
\end{eqnarray}
which becomes
\begin{eqnarray}
c=c_1+\frac{\tanh(\delta\lambda)}{2\delta\lambda}\Big(2\alpha(c_x^2+c_y^2+c_z^2)+c_z\log(\frac{a}{b})\Big),\nonumber
\end{eqnarray}
or
\begin{eqnarray}
c=c_1+\tanh\Big(\frac{1}{2}\sqrt{\Big(2\alpha c_z+\log(\frac{a}{b})\Big)^2+4\alpha^2(c_x^2+c_y^2)}\Big)\frac{2\alpha(c_x^2+c_y^2+c_z^2)+c_z\log(\frac{a}{b})}{\sqrt{\Big(2\alpha c_z+\log(\frac{a}{b})\Big)^2+4\alpha^2(c_x^2+c_y^2)}}.\nonumber\\\label{spinsol}
\end{eqnarray}
This equation is monotonic in $\alpha$ and therefore it is uniquely specified by the value of $c$. Ultimately this is a consequence from the concavity of the entropy. The proof of (\ref{spinsol})'s monotonicity is below:
\paragraph{Proof:}
For $\hat{\rho}$ to be Hermitian, $\hat{C}$ is Hermitian and $\delta\lambda=\frac{1}{2}\sqrt{f(\alpha)}$ is real. Further more, because $\delta\lambda$ is real $f(\alpha)\geq0$ and thus $\delta\lambda\geq0$. Because $f(\alpha)$ is quadratic in $\alpha$ and positive, it may be written in vertex form,
\begin{eqnarray}
f(\alpha)=a(\alpha-h)^2+k,
\end{eqnarray}
where $a>0$, $k\geq 0$, and $(h,k)$ are the $(x,y)$ coordinates of the minimum of $f(\alpha)$. Notice that the form of (\ref{spinsol}) is,
\begin{eqnarray}
F(\alpha)=\frac{\tanh(\frac{1}{2}\sqrt{f(\alpha)})}{\sqrt{f(\alpha)}}\times\frac{\d f(\alpha)}{\d\alpha}.
\end{eqnarray}
Making the change of variables $\alpha'=\alpha-h$ centers the function such that $f(\alpha')=f(-\alpha')$ is symmetric about $\alpha'=0$. We can then write,
\begin{eqnarray}
F(\alpha')=\frac{\tanh(\frac{1}{2}\sqrt{f(\alpha')})}{\sqrt{f(\alpha')}}\times 2a\alpha',
\end{eqnarray}
where the derivative has been computed. Because $f(\alpha')$ is a positive, symmetric, and monotonically increasing on the (symmetric) half-plane (for $\alpha'$ greater than or less that zero), $S(\alpha')\equiv\frac{\tanh(\frac{1}{2}\sqrt{f(\alpha')})}{\sqrt{f(\alpha')}}$ is also positive and symmetric, but it is unclear whether or not $S(\alpha)$ is also monotonic in the half-plane. We may restate
\begin{eqnarray}
F(\alpha')=S(\alpha')\times 2a\alpha'.
\end{eqnarray}
We are now in a decent position to preform the derivate test for monotonic functions:
\begin{eqnarray}
\frac{\d}{\d\alpha'}F(\alpha')&=&2aS(\alpha')+2a\alpha'\frac{\d}{\d\alpha'} S(\alpha')\nonumber\\
&=& 2aS(\alpha')\Big(1-\frac{a\alpha'^2}{a\alpha'^2+k}\Big)+a\frac{a\alpha'^2}{a\alpha'^2+k}\Bigg(1-\tanh^2(\frac{1}{2}\sqrt{a\alpha'^2+k})\Bigg)\nonumber\\
&\geq&2aS(\alpha')\Big(1-\frac{a(\alpha')^2}{a\alpha'^2+k}\Big)\geq0\nonumber\\
\end{eqnarray}
because $a,k,S(\alpha')$, and therefore $\frac{a\alpha'^2}{a\alpha'^2+k}$ are all $>0$.  The function of interest $F(\alpha')$ is therefore monotonic for all $\alpha'$, and therefore it is monotonic for all $\alpha$, completing the proof that  there exists a unique real Lagrange multiplier $\alpha$ in (\ref{spinsol}).

Although (\ref{spinsol}) is monotonic in $\alpha$ it is seemingly a transcendental equation. This can be solved graphically for the given values $c,c_1,c_x,c_y,c_z$, i.e. given the Hermitian matrix and its expectation value are specified. Equation (\ref{spinsol}) and the eigenvalues take a simpler form when $a=b=\frac{1}{2}$, because in this instance $\hat{\varphi}\propto\hat{1}$ and commutes universally so it may be factored out of the exponential in (\ref{spinrhoexample}).

\end{document}